# Bimetal-and-electret-based thermal energy harvesters – Application to a battery-free Wireless Sensor Node

S. Boisseau[1], P. Gasnier[1], S. Monfray[2], G. Despesse[1], O. Puscasu[2], A. Arnaud[2], T. Skotnicki[2]

[1] CEA, Leti, Minatec Campus, 17 rue des Martyrs - 38054 Grenoble Cedex 9, France
[2] STMicroelectronics, 850 rue J. Monnet - 38926 Crolles, France

E-mail: sebastien.boisseau@cea.fr

**Abstract.** This paper introduces a thermal energy harvester turning thermal gradients into electricity by coupling a bimetallic strip to an electret-based converter: the bimetallic strip behaves as a thermal-to-mechanical power converter turning thermal gradients into mechanical oscillations that are finally converted into electricity with the electret. Output powers of 5.4µW were reached on a hot source at 70°C, and, contrary to the previous proofs of concept, the new devices presented in this paper do not require forced convection to work, making them compatible with standard conditions of thermal energy harvesting and environments such as hot pipes, pumps and more generally industrial equipment. Finally, ten energy harvesters have been parallelized and combined to a self-starting power management circuit made of a flyback converter to supply a battery-free Wireless Temperature Sensor Node, sending information every 100 seconds after its startup state.

**Keywords.** Bimetal, bimetallic strip, electret, thermal energy harvesting, Wireless Sensor Networks, battery-free devices

## 1. Introduction

Thermal power is lost in many systems: motors, pumps, hot pipes, electrical distribution, electronic components, etc. This power can be exploited to supply small and low-consumption devices such as Wireless Sensor Nodes, giving them a theoretical unlimited lifetime and removing any maintenance issues such as battery replacement or recharging. These fully-autonomous Wireless Sensor Nodes gathered in networks and powered by the surrounding thermal energy can be of great benefit for industrial equipment, enabling to implement predictive maintenance and to prevent unexpected failures by exploiting the natural heating of machines.

Thermoelectric energy harvesters, based on Seebeck effect, are today the most widely used devices to turn thermal power into electricity for Wireless Sensor Nodes (MicroPelt, EnOcean, Nextreme). Yet, to be efficient, these devices require quite expensive materials such as Bismuth Telluride ($Bi_2Te_3$).

The thermal energy harvesters (TEH) presented in this paper can be an alternative to these thermoelectric energy harvesters. Based on a bimetallic strip to turn thermal gradients into a mechanical oscillation and an electret to convert this movement into electricity, bimetal-and-electret-based TEH can be made from simple and fully available materials, reducing costs and facilitating mass production. In fact, first proofs of concept have already been presented in [1] with a piezoelectric conversion, and in [2] with this electret-based approach. Yet, the first harvesters presented in [2] required forced convection to work, which is not compatible with the concept of ambient thermal energy harvesting. Therefore, the scavengers have been improved and this drawback has been removed. The new design of bimetal-and-electret-based TEH is introduced in section 2. Section 3 presents the output powers as a function of hot sources temperatures and electrets surface voltages, showing a high electromechanical coupling of the electrostatic converter. A self-starting power management circuit is introduced in section 4 and combined to 10 TEH in parallel. Finally, the implementation of a battery-free Wireless

Temperature Sensor Node powered by these new energy scavengers definitely validates the proof of concept.

## 2. Bimetal-and-electret-based thermal energy harvesters – Improvements and compatibility with ambient thermal conditions

Bimetal-and-electret-based TEH are a combination of curved bimetallic strips (thermal-to-mechanical converters) and electret-based converters (mechanical-to-electrical transducers) whose operation principles are presented hereafter.

### 2.1 Curved bimetallic strips and thermal-to-mechanical power conversion

Bimetals are made of two different metals with different coefficients of thermal expansion (CTE) that are joined together. Thanks to the CTE difference, flat bimetallic strips bend when heated up or cooled down (Figure 1a). Curved and stamped bimetallic strips (Figure 1b) have strong nonlinear behaviors and may be able to suddenly snap and snap-back (sudden buckling) between two positions according to the temperature with a hysteretic behavior (figure 1c) [3-6]. Thanks to these properties, flat, curved and stamped bimetallic strips are today used as thermal actuators in many devices such as switches, clocks, thermostats, thermometers, irons, etc [7-9].

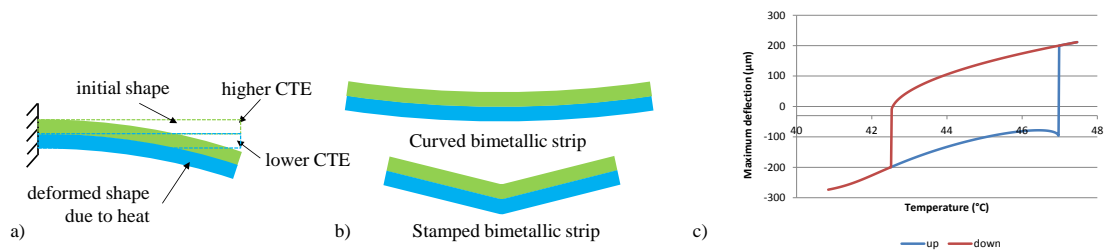

Figure 1. (a) Flat bimetallic strip, (b) curved and stamped bimetallic strips and (c) hysteresis cycle

Here, curved bimetallic strips are used as heat engines converting the thermal energy into a mechanical work (figure 2). A bimetallic strip is inserted in a cavity whose lower plate is hot and whose upper plate is cold. The cycle starts when the system is in its "lower state" (figure 2a). The bimetallic strip is in contact with the lower plate and its temperature is below the snapping temperature. As it is in contact with the lower plate, the bimetallic strip heats up, accumulates mechanical elastic energy until its temperature reaches the snapping temperature ($T_s$). Then, it suddenly snaps to its upper state and enters in contact with the upper plate (figure 2b) where it is cooled down. When the bimetallic strip's temperature reaches the snapping-back temperature ($T_{sb}$), it suddenly snaps-back to its lower state and a new cycle restarts.

As a consequence, the thermal power is converted into a mechanical oscillation of the bimetallic strip provided that (i) the snapping temperature ($T_s$) is lower than the lower plate's temperature ($T_{hot}$) and (ii) the snapping-back temperature ($T_{sb}$) is higher than the upper plate's temperature ($T_{cold}$).

$$T_{hot} > T_s > T_{sb} > T_{cold} \qquad (1)$$

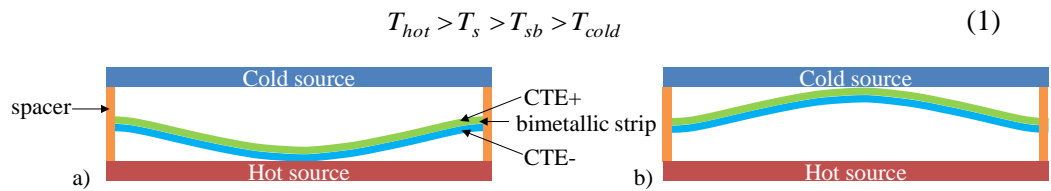

Figure 2 – Bimetal-based thermal-to-mechanical power converter (a) lower state and (b) upper state

The mechanical oscillation can be then converted into electricity thanks to a mechanical-to-electrical power converter such as a piezoelectric material [1,10-11], a coil-magnet architecture or an electrostatic converter, as presented hereafter.

## 2.2 Electret-based converters and mechanical-to-electrical power conversion

Electrets are electrically charged dielectrics able to keep their charges for years. They can be considered as the equivalent of magnets in electrostatics. Many materials have shown electret properties; the most common are: Teflon® [12-13], $SiO_2/Si_3N_4$/HMDS [14] and CYTOP® [15]. The first application of electrets was microphones, but, they have been increasingly employed in electrostatic energy harvesters since the 2000s as a permanent polarization source enabling a direct mechanical-to-electrical power conversion [16-20].

The basic electret-based electrostatic converter is made of two plates forming a capacitor (figure 3a). An electret layer is added on one (or both) of the two plates. The charge of the electret is constant and equal to $Q_i$. According to Gauss's law, charges appear on the two electrodes ($Q_1$ on the lower plate and $Q_2$ on the upper plate) so that the sum of charges $Q_1+Q_2$ is equal to $Q_i$. A movement of the upper plate relative to the lower one induces a movement of charges between the two plates due to a variation of the electrostatic induction. Then, the mechanical movement of the upper plate is turned into a circulation of charges through the load: mechanical power is converted into electricity.

The equivalent model of the electret-based electrostatic converter is presented in figure 3b. It is ruled by (2) and (3) [12] where $V_s$ is the surface voltage of the electret, $C(t)$ the capacitance between the two plates, $R$ the load and $p(t)$ the instantaneous output power.

$$\frac{dQ_2}{dt} = \frac{V_s}{R} - \frac{Q_2}{C(t)R} \quad (2)$$

$$p(t) = R\left(\frac{dQ_2}{dt}\right)^2 \quad (3)$$

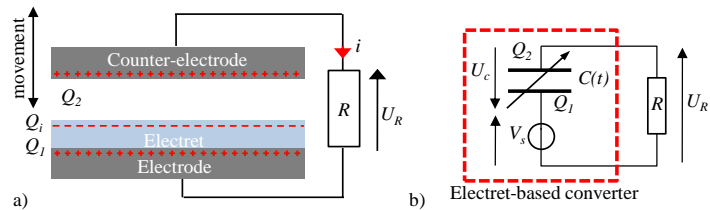

Figure 3 – (a) Electret-based electrostatic converter and (b) equivalent model

## 2.3 Bimetal-and-electret-based thermal energy harvesters with shifted hot sources

The electret-based electrostatic converter is combined to the above-mentioned bimetal-based thermal-to-mechanical transducer by adding an electret layer on one or both of the plates of the heat engine, as depicted in figure 4a. Due to the thermal gradient between the two plates, the bimetallic strip oscillates between its lower state (figure 4a) and its upper state (figure 4b). This oscillation is turned into electricity by the electrostatic converter, and globally, thermal power is turned into electricity.

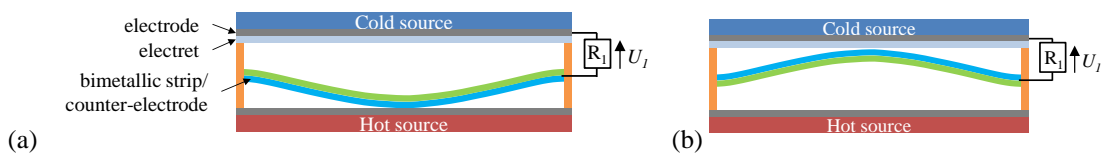

Figure 4 – Bimetal-and-electret-based converter (a) lower state and (b) upper state with one electret-based converter on the upper plate

For the first proofs of concept [2], thin steel layers covered by electrets were placed around the curved bimetallic strip. When set on a hot source at 60°C, the bimetallic strip oscillated between its two states (figures 5a and 5b), and prototypes (figure 5c) have shown output powers in the 10µW range.

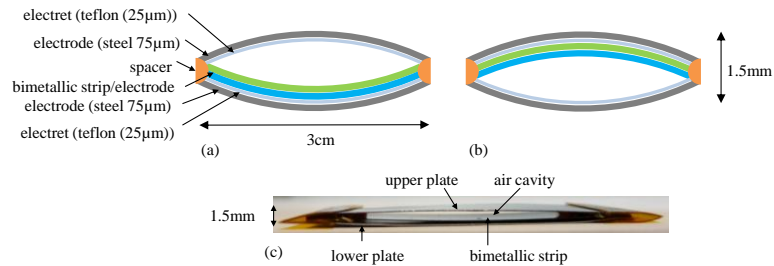

Figure 5 – First proofs of concept (a) lower state, (b) upper state and (c) prototype

Yet, due to non-optimized thermal control between the two steel layers, it was not possible to maintain a sufficient gradient that would keep the bimetallic strip oscillating, without forced convection. And as a consequence, it was hard to find environments that would be compatible with these energy harvesters as both a hot source and forced convection were required. To overcome this drawback, the design has been rethought with the idea of limiting thermal bridges between the two plates.

However, to oscillate, the bimetallic strip needs to alternatively touch the hot and the cold source. And, as the peak-to-peak deflection of the bimetallic strip is in the order of 500µm, the hot and the cold source must be separated by 500µm maximum. The solution that was implemented consists in bringing the hot source to the center of the bimetallic strip by a small cylinder of copper (figure 6). A heat sink has also been added on the upper plate to further increase the thermal gradient between the two plates and the heat flux in the energy harvester. And, contrary to the first devices presented in [2], only one electret-based converter is used and added on the upper plate.

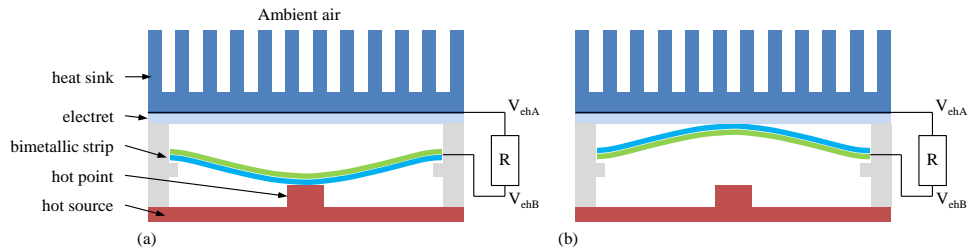

Figure 6 – Improved design with a shifted hot source and a heat sink (a) lower state and (b) upper state

With the heat sink and the cylinder of copper, the devices now work only with a hot source making them compatible with standard WSN environments (motors, pumps, industrial equipment…).

## 3. Energy Harvesters, output voltages and output powers

### 3.1 Prototypes

A prototype of the new bimetal-and-electret-based TEH made of a plastic support to limit heat conduction, a curved bimetallic strip, an electret, a cylinder of copper and a heat sink is presented in figure 7.

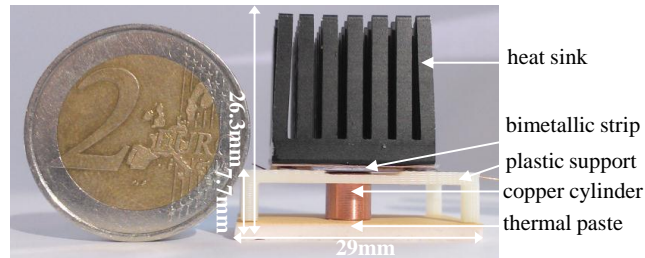

Figure 7 – Prototype

The bimetallic strips are in Invar (Fe-Ni36) whose CTE=$2\times10^{-6}$ and in B72M (Mn-Cu18-Ni10) whose CTE=$26.4\times10^{-6}$ and size 2cm×1cm×0.115mm. They are stamped according to an elliptic paraboloid shape and designed to snap at $T_s$=47°C and to snap back at $T_{sb}$=43°C. The shapes of the bimetallic strips before and after snapping are respectively presented in figure 8a and figure 8b. They are finally covered by a 1µm-thick parylene-C layer to avoid electret-metal contacts during oscillations.

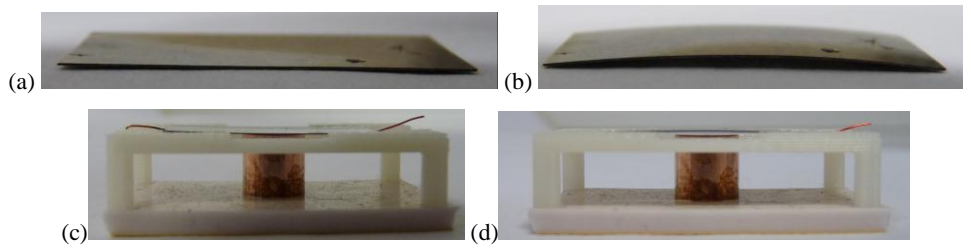

Figure 8 – Bimetallic strip (a) before snapping, (b) after snapping. Bimetallic strip in the structure (c) before snapping and (d) after snapping

The heat sink has a surface of 20mm×20mm, a height of 19.1mm and an absolute thermal resistance of $R_{th}$=14K/W. The cylinder of copper has a diameter of 6mm and a height of 6mm. As in [2], 25µm-thick Teflon layers have been chosen for the electrets. The sheets have been metalized on the rear face and glued on a sheet of copper and finally glued on the heat sink. They have been charged by a negative triode corona discharge with a point voltage of 10kV for 1h. During the first 30 min of the corona charging process, the electret was heated at 200°C on a hotplate. Then, the hotplate was turned off and the corona discharge was maintained while the electret and the hotplate cooled down. This process enables to make stable Teflon electrets even with elevated temperatures and contacts. Besides, the stability of these electrets has already been validated up to 500V in [2] for 850'000 cycles.

The copper cylinder is inserted in the plastic structure before adding the bimetallic strip. Finally, the heat sink covered by the electret is placed on the structure presented in figure 8c (lower state) and 8d (upper state). The electronic circuit is connected between the bimetallic strip and the sheet of copper, and the TEH is finally set on the hot source (60-85°C).

### 3.2 Output voltages and output powers

Three examples of output voltages on a load R=1GΩ for $T_{hot}$=75°C and for various electret's surface voltage are presented in figure 9 (cooling by ambient air). Because of the electret-based converter, the output voltages of the energy harvester reach hundreds of volts. It is also interesting to note the impact of the electrostatic converter on the bimetallic strip's snapping frequency: the higher the electret's surface voltage, the higher the output power and the lower the snapping frequency. This strong impact of the electret's surface voltage on the mechanical behavior of the TEH proves the high electromechanical coupling of the electrostatic converter. These results are gathered in figure 10a presenting the output powers and the snapping frequencies as a function of $V_s$. It is also noteworthy that the output power as a function of the electret's surface voltage saturates at 6.4µW, and increasing $V_s$ above 600V has no benefit. In fact, increasing $V_s$ increases the output energy per cycle but decreases the snapping frequency and finally, the output power remains the same (figure 10a).

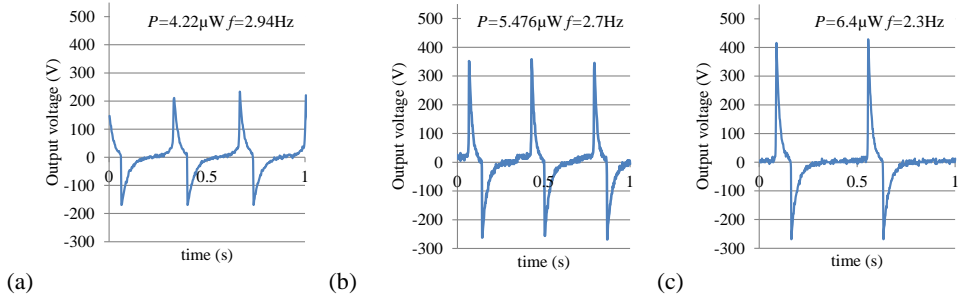

Figure 9 – Output voltages ($U_R$) and output power for R=1GΩ, $T_{hot}$=75°C and for (a) $V_s$=375V, (b) $V_s$=500V and (c) $V_s$=875V (cooled by ambient air)

Moreover, since these thermal energy harvesters use bimetallic strips as thermal-to-mechanical converters, their operating temperatures are limited (1). Figure 10b shows the output power and the snapping frequency as a function of the hot source temperature for $V_s$=500V. It demonstrates that the device works for $T_{hot}$ comprised between 63°C and 83°C. However, snapping frequencies and output powers are highly reduced outside the 70°C-75°C optimum temperature range.

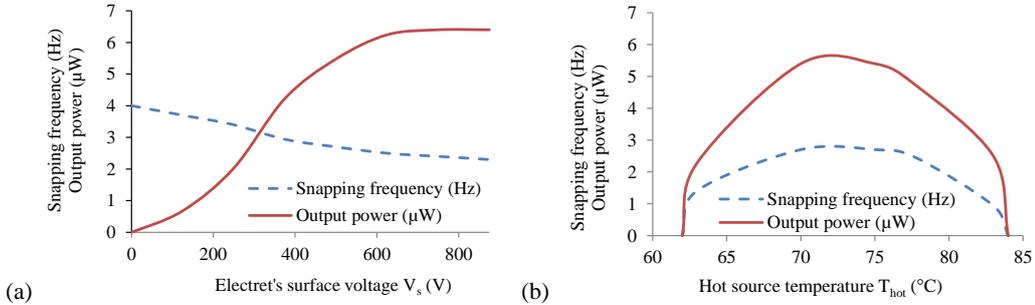

Figure 10 – (a) Snapping frequency and output power as a function of the electret's surface voltage for $T_{hot}$=75°C (cooled by ambient air) and (b) Snapping frequency and output power as a function of the temperature of the hot source for $V_s$=500V (cooled by ambient air).

Bimetal-and-electret-based TEH have a limited working temperature range. Then, to harvest power from various temperature conditions, devices with different characteristics are necessary; this can be easily obtained by changing the bimetallic strip or by modifying the size of the copper cylinder or the heat sink.

## 4. Power management circuit, self-startup and application

As presented in figure 9, the output voltages of these energy scavengers are alternative and can reach some hundreds of volts, which is not compatible with standard electronic circuits such as sensors or microcontrollers. As a consequence, a power management circuit has to be implemented to convert the TEH output into a 3V-DC supply source. Furthermore, as the output power of one TEH is not sufficient to power a Wireless Sensor Node, parallelization is required.

### 4.1 Energy harvesting from multiple devices in parallel

Like in [2], a power transfer on maximum voltage detection (SECE [21]) has been chosen to transfer the energy from the energy scavengers in parallel into a buffer through a flyback converter (figure 9a). The power management circuit is made of a magnetic circuit ($L_p$-$L_s$), two controlled transistors ($K_p$ and $K_s$), a Control Circuit which commands them and a buffer capacitor ($C_b$) as presented in figure 11a. The theoretical voltages and currents in the flyback converter during the energy transfer are depicted in figure 11b. $U_{EH}$ and $U_{Cb}$ are respectively the voltages across the energy harvesters in parallel and the buffer; $i_p$ and $i_s$ are the currents flowing through the primary and the secondary windings. When one of the energy harvesters' output voltage reaches its maximum, the Control Circuit closes $K_p$ during $T_1$ to transfer the energy $E=\frac{1}{2}C_{max}U_{max}^2$ stored in its capacitor into the magnetic circuit: $U_{EH}$ drops to 0 while $U_{Cb}$ stays constant. $K_p$ is then opened, and $K_s$ is closed to transfer $E$ into the buffer capacitor: $U_{EH}$ stays equal to 0 and $U_{Cb}$ increases from $U_{Cb}'$ to $U_{Cb}''$. Thanks to this circuit, the most part of the energy stored in the energy harvester's capacitance is transferred to $C_b$.

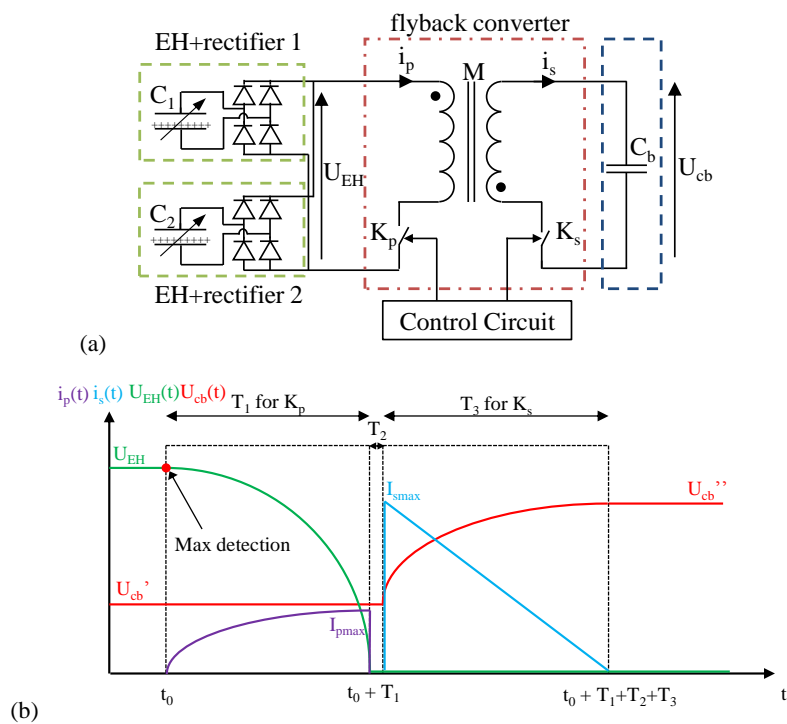

Figure 11 – (a) Parallelization of devices and power management circuit and (b) theoretical voltages and currents in the flyback converter during an energy transfer

For this preliminary measure, the power management circuit is placed at the output of one TEH and powered by an external DC supply source. The electronic components used in the flyback converter are overviewed in table 1 at the end of the paper. As expected, the Control Circuit detects the output voltage's maxima (figure 12a) and $U_{EH}$, $U_{Cb}$ and $i_s$ during an energy transfer (figure 12b) are in agreement with the theoretical behavior of flyback converters. Yet, the output power is highly limited due to the parasitic capacitances induced by the diode bridge, the windings of the magnetic circuit and the transistors. At the end of the flyback conversion process, an output power of about 1.4µW per device is achieved. It corresponds to a global efficiency of 20%-25%: 50% of the TEH output power is lost in the parasitic capacitances and the flyback converter's efficiency is 50%.

With ten devices in parallel, an output power comprised between 5 and 10µW is harvested. It is not equal to ten times the output power of one device for two reasons:

(i) TEH and their output are not perfectly reproducible for this first demonstration. Some devices harvest less power than the one presented in section 3.

(ii) The capacitances of the TEH in parallel with a scavenger act like parasitic capacitances and reduce its output power. This effect could be reduced by increasing the variation of capacitance between the two states, for example by using thinner electret layers or conformable electrodes.

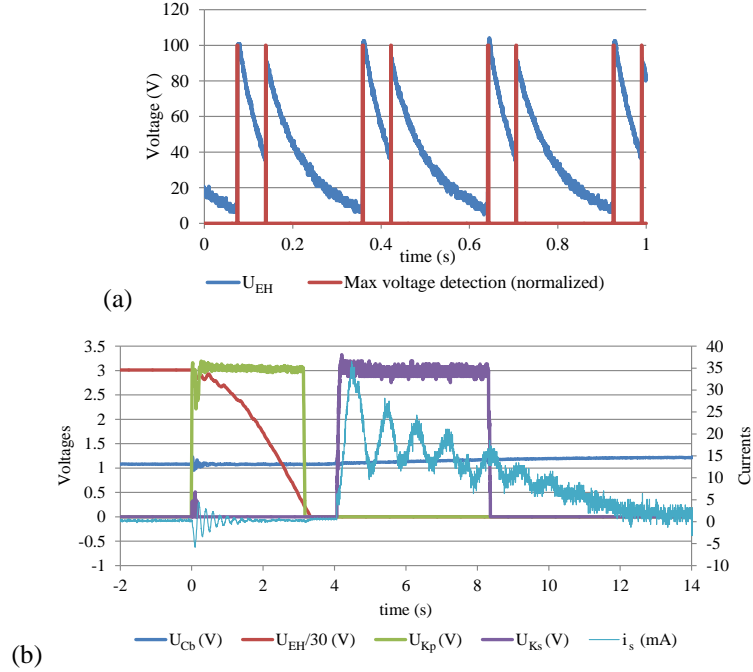

Figure 12 – (a) Maximum voltage detection, (b) zoom on an energy transfer

The Control Circuit that detects the output voltages' maxima and generates $T_1$ and $T_3$, which controls $K_p$ and $K_s$, consumes between 500nA and 1µA@3V (1.5-3µW) depending on the maximum voltage detection frequency. In [2], this circuit was powered by an external supply source (battery or stabilized power supply).

In this paper, we propose a major improvement which enables to start and to power the Control Circuit without any initial energy with the aim of developing battery-free Wireless Sensor Nodes. Battery-free devices can be of great benefit for environmental, lifetime or harsh environments compatibilities reasons. However, supplying the above-mentioned power management circuit from scratch is not an easy task as, at the beginning, without energy, the Control Circuit cannot be powered and the transistors cannot be controlled.

### 4.2 Self-startup

The solution we propose consists in starting the Control Circuit by accumulating energy in a buffer ($C_s$) through the diode bridges ($D_B$) before switching to the flyback conversion process when the voltage across $C_s$ ($U_{Cs}$) is sufficient to power the Control Circuit. The whole architecture is depicted in figure 13a; no external supply source is added.

It is composed of the previous power management circuit, a depletion-mode MOSFET ($K_{bp}$) which bypasses the flyback converter, a pMOSFET ($K_{cc}$) which cuts the power supply of the Control Circuit and two buffer capacitors $C_b$ and $C_s$ connected by a diode $D_C$. $K_{bp}$ and $K_{cc}$ are controlled by the "Startup Control" circuit which is a Schmitt trigger, observing $U_{Cs}$ and switching at $U_{Cs}^+$ and $U_{Cs}^-$.

Four states can be identified during the startup process (figure 13b):

*State 0 – no-energy state.* At the beginning, there is no energy in $C_b$ and $C_s$ ($U_{Cb}=0$ and $U_{Cs}=0$). The depletion-mode MOSFET $K_{bp}$ is closed (its gate-source voltage is equal to 0) and $K_{cc}$ is open.

*State 1 – non-optimized path: diode-bridge-capacitor conversion.* The energy scavengers start to harvest energy. As $K_{bp}$ is closed and $K_{cc}$ is open, the flyback converter is bypassed by the depletion-mode MOSFET $K_{bp}$. The power management circuit behaves like a simple diode-bridge-capacitor circuit and the energy goes directly to $C_s$ via the diode bridges $D_B$; $U_{Cs}$ increases.

*State 2 – optimized path: flyback conversion.* When $U_{Cs}$ reaches $U_{Cs}^+$, the Startup Control circuit opens $K_{bp}$ and closes $K_{cc}$. The Control Circuit is supplied by $C_s$ and the optimized conversion process through the flyback starts; $U_{Cb}$ increases.

*Return to state 1.* As the Control Circuit consumes energy and since $C_s$ does not receive energy from the TEH anymore ($K_{bp}$ is open), $U_{Cs}$ decreases. When $U_{Cs}$ reaches $U_{Cs}^-$, the startup circuit closes $K_{bp}$ and opens $K_{cc}$. The power management circuit behaves again like a diode-bridge-capacitor system (state 1), to recharge $C_s$.

*Return to state 2.* When $U_{Cs}$ reaches $U_{Cs}^+$ again, the conversion process through the flyback restarts (state 2). The switching between state 1 and state 2 continues until $U_{Cb}$ reaches $U_{Cs}+V_{DCth}$, where $V_{DCth}$ is the threshold voltage of diode $D_C$.

*State 3 – end of startup.* The power management circuit remains in the flyback conversion configuration, optimizing the power extraction from the energy harvesters. $K_{cc}$ stays closed, $K_{bp}$ open.

The energy stored in $C_b$ can be finally used to power a Wireless Sensor Node (figure 13b).

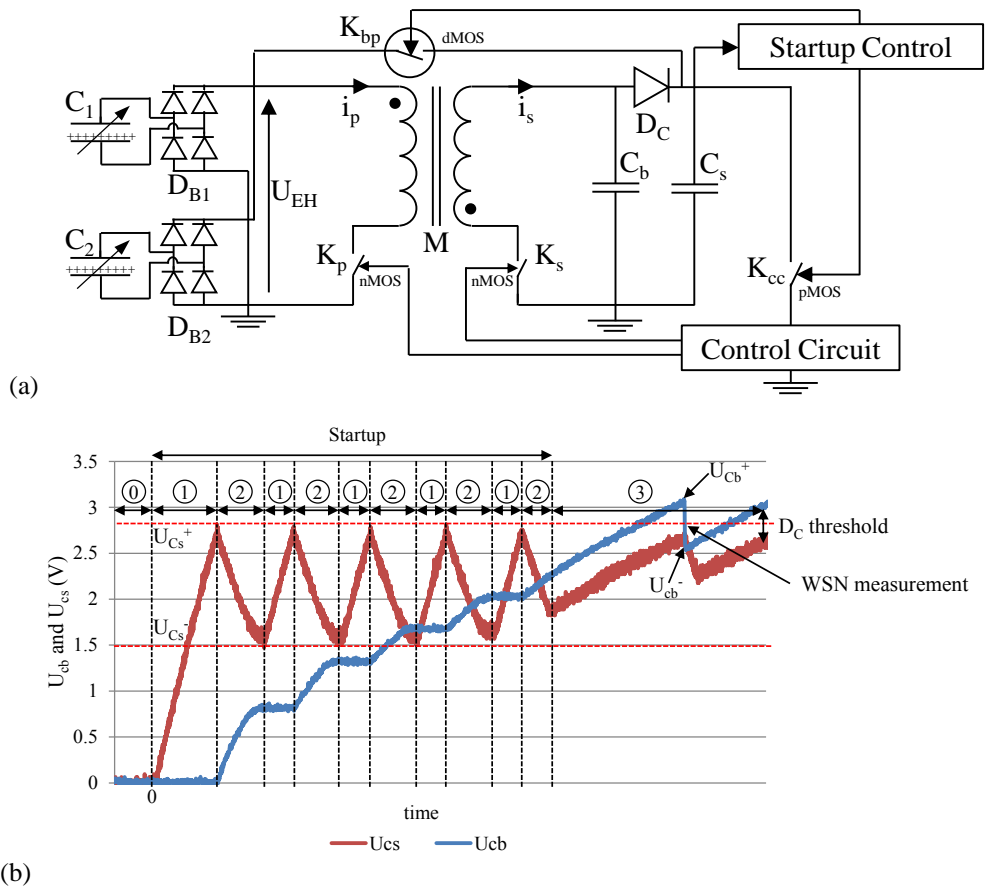

Figure 13 – (a) Power management circuit with self-startup and (b) $U_{Cb}$ and $U_{Cs}$ as a function of the time showing the various states

The self-starting power management circuit is combined to 10 bimetal-and-electret-based thermal energy harvesters in parallel, producing between 5 and 10μW together, to power a wireless temperature sensor node.

### 4.3 Application - A battery-free Wireless Sensor Node powered by bimetal-and-electret-based thermal energy harvesters

A wireless sensor node is connected to $C_b$ with a pMOSFET ($K_{app}$). $K_{app}$ is commanded by the Application Control which is a Schmitt trigger observing $U_{Cb}$ and switching at $U_{Cb}^+$ and $U_{Cb}^-$.

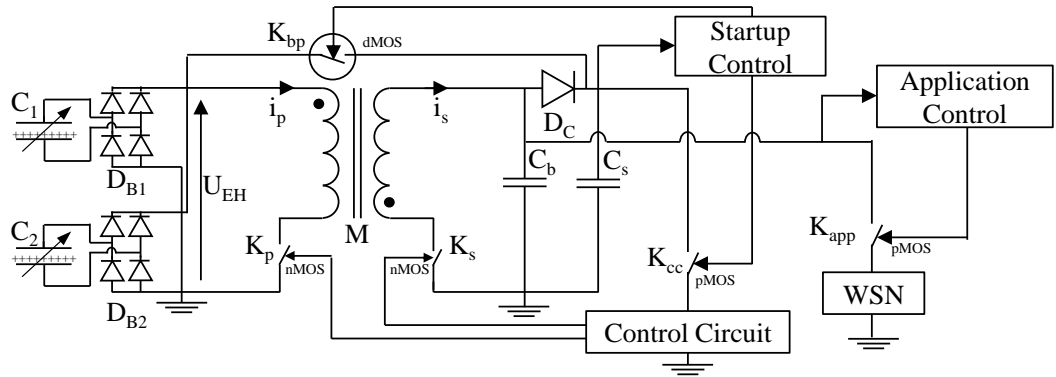

Figure 14 – Wireless Temperature Sensor Node connected to the power management circuit

At the beginning, $K_{app}$ is open and the sensor node (WSN) is not powered. As soon as $U_{Cb}$ reaches $U_{Cb}^+$, the Schmitt trigger closes $K_{app}$. The sensor node is powered, starts, performs a series of temperature measurements and data emissions until $U_{Cb}$ reaches $U_{Cb}^-$. The Application Control opens $K_{app}$, disconnecting the sensor node from $C_b$, waiting $U_{Cb}$ to reach $U_{Cb}^+$ again to start a new measurement cycle.

The wireless sensor node, emitting at 868MHz, has been programmed to make a temperature measurement with an external temperature sensor and to send it in RF with its emitter. The energy consumption of each measurement of the WSN (waking up, measure, emission) is about 100µJ.

$U_{Cb}$, $U_{Cs}$ and $U_{WSN}$ are presented in figure 15 from the zero-energy state to WSN data emissions. First of all, it is noteworthy that the startup of the power management circuit takes quite a long time (about 1000s), essentially due to the low efficiency of the diode-bridge capacitor conversion. But, as soon as the startup state is over, a complete measurement can be performed every 100s with 10 bimetallic strips, each sudden drop of $U_{Cb}$ in figure 15 corresponding to a complete measurement including the RF emission. This validates the suitability of bimetal-and-electret-based TEH for Wireless Sensor Nodes.

Moreover, the durations (startup, between RF emissions) are already compatible with many needs in predictive maintenance. Obviously, they could be reduced by using more bimetallic strips in parallel and the startup time could also be totally removed by employing a storage battery.

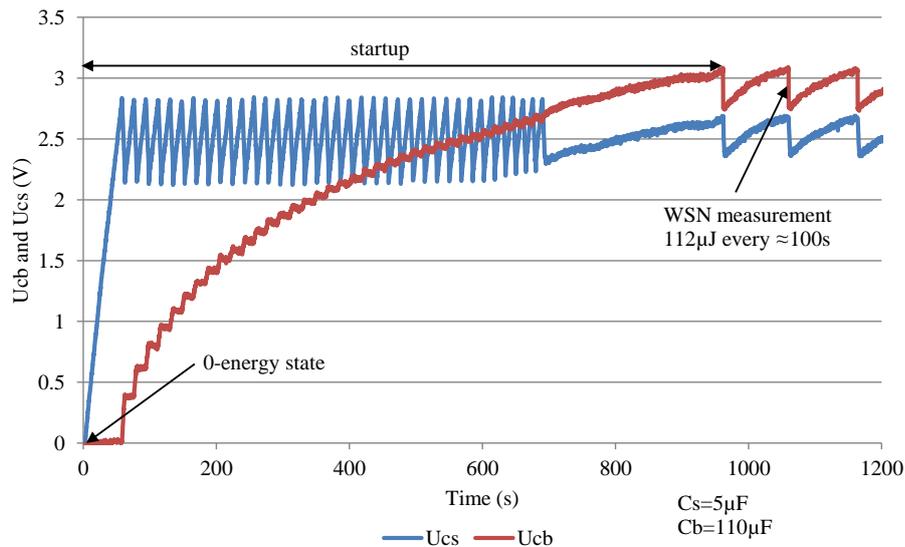

Figure 15 – $U_{Cb}$, $U_{Cs}$ and $U_{WSN}$ as a function of the time

Table 1 – Electronic components used in this paper

| Component | Value/Reference | Component | Value/Reference |
|---|---|---|---|
| $D_B$ | MMBD1503 | $L_p$ | 98.2mH |
| $D_C$ | MMBD1503 | $L_s$ | 85μH |
| $K_p$ | BSP130 | Magnetic core | 3F3 – EFD20 |
| $K_s$ | FDS9926 | $C_b$ | 110μF |
| $K_{cc}$ | FDN306P | $C_s$ | 5μF |
| $K_{app}$ | FDN306P | $K_{bp}$ | BSS139 |

## 5. Conclusions and Perspectives

We have presented new bimetal-and-electret-based thermal energy harvesters turning thermal gradients into electricity in a two-step conversion. The use of a heat sink and a cylinder of copper have enabled to make the energy harvesters work without forced convection. We managed to harvest up to 5.4μW per device on a hot source at 70°C and the high electromechanical coupling induced by the electret-based electrostatic converter has been proven. A power management circuit able to start from scratch has been developed by exploiting a passive diode-bridge-capacitor path and an active and optimized flyback conversion process. Again, the strong impact of parasitic capacitances has been raised and the global efficiency of the flyback power converter is only 20%-25% for this reason. Finally, bimetal-and-electret-based TEH have been combined to a wireless temperature sensor node. A complete measurement is performed each 100 seconds with 10 devices in parallel, definitely validating their viability for WSNs.

Research is now focused (i) on the size reduction of the bimetallic strips by using cleanroom processes to increase power densities, (ii) on thermal management optimizations to remove the heat sink and to reduce the energy harvesters' thickness (already validated with piezoelectric devices [1]) and (iii) on a mean to reduce the parasitic capacitances induced by the power management circuit.

 Acknowledgments

The authors would like to thank G. Pitone and G. Delepierre, Delta Concept, Sassenage, France for the design of the bimetallic strips; and M. Gallardo, F. Pfister and S. Brulais for the electronic circuits and the microcontroller programs.

This work has been partially funded by the European Regional Development Fund (FEDER/ERDF) and by the French Inter-ministerial Fund (FUI), through HEATec project.